\documentclass[pra,twocolumn,showpacs,twoside]{revtex4}

\usepackage{amssymb, amsbsy, amsmath, latexsym, dsfont, array, layout, graphicx}
\usepackage{color}

\newcommand{\ket}[1]{\left|{#1}\right\rangle}
\newcommand{\bra}[1]{\left\langle{#1}\right|}

\begin{document}

\title{Universal quantum computing with nanowire double quantum
dots}
\author{P. Xue}
\affiliation{Department of Physics, Southeast University, Nanjing
211189, P. R. China}
\date{\today}

\begin{abstract}
We show a method for implementing universal quantum computing using
of a singlet and triplets of nanowire double quantum dots coupled to
a one-dimensional transmission line resonator. This method is
attractive for both quantum computing and quantum control with
inhibition of spontaneous emission, enhanced spin qubit lifetime,
strong coupling and quantum nondemolition measurements of spin
qubits. We analyze the performance and stability of all required
operations and emphasize that all techniques are feasible with
current experimental technology.
\end{abstract}
\pacs{03.67.Mn, 42.50.Pq, 73.21.La, 03.67.Lx}
\maketitle
\section{Introduction}

A quantum computer comprising many two-level systems---qubits,
exhibits coherent superpositions and entanglement. Quantum
computing, which is based on these features, enables some
computational problems to be solved faster than would ever be
possible with a classical computer~\cite{Gro97}, and exponentially
speeds up solutions to other problems over the best known classical
algorithms~\cite{Sho94}, is currently attracting enormous interests.
Among the promising candidates for quantum computing, solid-state
implementations such as spin qubits in quantum dots~\cite{LD97} and
bulk silicon~\cite{Kan98}, and charge qubits in bulk
silicon~\cite{ABW+07} and in superconducting Josephson
junctions~\cite{WSB+05}, are especially attractive because of
stability and expected scalability of solid-state systems; of these
competing technologies, semiconductor double quantum dots (DQDs) are
particularly important because of the combination spin and charge
manipulations to take advantage of long memory times associated with
spin states and at the same time to enable efficient readout and
coherent manipulation of charge states.

Our goal is to develop a realizable architecture for semiconductor
quantum computation. The qubit is manifested as a nanowire (NW)
quantum dot pair such that each having an electron and thus the
singlet and one of the triplets of two-electron states correspond to
the logical state $\ket{0}$ and the orthogonal state $\ket{1}$. The
resonator-assisted interaction between DQDs and a microwave
transmission line resonator (TLR) is used to implement a universal
set of quantum gates and readout of the qubits.

From two points, we show the advantages of our scheme compared to
the previous proposals on semiconductor quantum computation.
Firstly, for the previous proposals which make use of single or
double quantum dots defined by a two-dimensional electron gas
(2DEG)~\cite{Im99,Petta05,Johnson05,Taylor07,GI06,Taylor06,Guo09,Xue10},
it would be difficult to implement a double-dot in a planar
resonator with lateral dots, shaped in a 2DEG by surface gates. The
reason being is that it would be difficult to prevent absorption of
microwaves in the 2DEG unless one can make the electric field
non-zero only in the double-dot region, which is not realistic
experimentally yet. Our strategy is to use the two-electron states
of DQDs inside NWs instead of 2DEG, which is more realistic for
implementing quantum computing experimentally.

Secondly, there are previous proposals making use of NW
DQDs~\cite{Trif08}, in which the spin-orbit interaction is used to
couple the spins and resonator. However the weak coupling between
the DQD spins and resonator mode is a challenge in experiments. Our
strategy to enhance the interaction is to make use of the coupling
between the electric dipole of charge states of DQDs and resonator,
which is much stronger compared to that in~\cite{Trif08}. However
the decoherence of charge states is another obstacle. In this work,
combining the advantages of spin and charge states and avoiding the
weak points of both, we propose a different mechanism, namely via
resonator-assisted interaction which leads to a strong coupling
between the resonator photons and effective electric dipole of the
state $\ket{0}$ and an ancillary state of DQDs, while the state
$\ket{1}$ is driven by a classical field, and then eventually
implements quantum control on the singlet and one of the triplet
spin states. Thus we encode the quantum information in spin states
and the quantum control is implemented via the charge dipole
transition which is driven by a TLR.

\begin{figure}[tbp]
   \includegraphics[width=8.5cm]{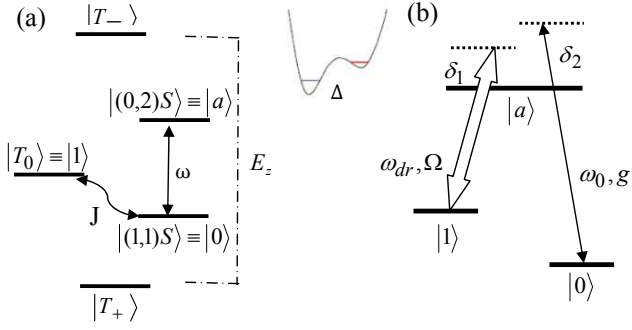}
  \caption{(a) Energy level diagram showing the $(0,2)$ and $(1,1)$ singlets, the three $(1,1)$ triplets
   and qubit states $\ket{(1,1)S}$ and $\ket{T_0}$ with the energy gap $J$ (the exchange energy $\sim T/10$).
   Schematic of the double-well potential with an energy offset $\Delta$ provided by the external
electric field. (b) The relevant three-level structure of DQDs. The
dipole transition $\ket{0}\rightarrow\ket{a}$ is coupled to the
fundamental mode of the resonator with the coupling coefficient $g$
and detuning $\delta_2$, while the transition
$\ket{1}\rightarrow\ket{a}$ is driven by a classical field with the
Rabi frequency $\Omega$ and detuning $\delta_1$.}
   \label{fig:energylevel}
\end{figure}

A solid-state realization cavity QED is proposed in Sec.~II and then
we discuss the case where the resonator and qubit are tuned on- and
off-resonance which can be used to implement a universal set of
gates including single- and two-qubit gates in Sec.~III A-C. The
initialization of qubit states can be implemented by an adiabatic
passage shown in Sec.~III D. The readout of qubits can be realized
via microwave irradiation of the TLR by probing the transmitted or
reflected photons shown in Sec.~III E. In Sec.~III F, the main
decoherence processes are dissipation of the TLR, charge-based
relaxation and dephasing of the NW DQDs occurring during gate
operations and transportation of qubits, and spin dephasing limited
by hyperfine interactions with nuclei. By numerical analysis we show
all gate operations and measurements can be implemented within the
coherent life time of qubits. Thus we address all Divincenzo
criteria~\cite{DiVincenzo} and show all play important roles in the
dynamics of the two-electron system but none represents a
fundamental limit for quantum computing. We make a summary in
Sec.~IV.

\section{System: Solid-state realization of cavity QED}
\subsection{The Hamiltonian}

We consider the system with two electrons located in adjacent
quantum dots coupling via tunneling. Imagine one of the dots is
capacitively coupled to a TLR~\cite{GI06,Taylor06,Guo09,Xue10}. We
assume that the left dot (L) is red-shifted with respect to the
right dot (R) and that the lowest conduction level of the left dot
is detuned by $\Delta$ with respect to the right one.

\begin{figure}[tbp]
   \includegraphics[width=8.5cm]{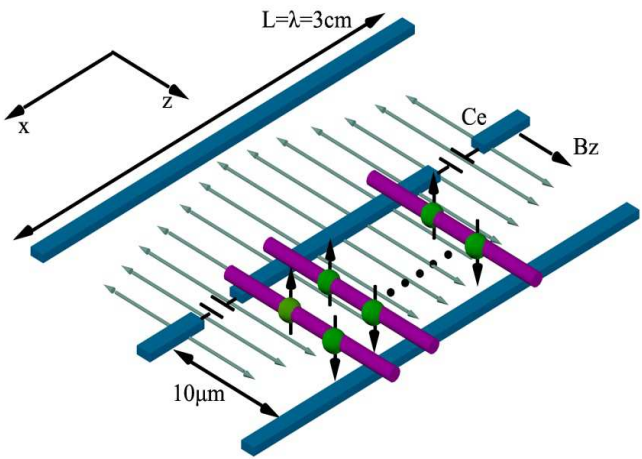}
  \caption{Schematic of
   NW DQDs capacitively coupled to the TLR. The coupling can be switched on and off via the external electric field.
   The DQD confinement can be achieved by barrier materials or by external gates (not shown).}
  \label{fig:schematic}
\end{figure}

In the $(1,1)$ regime, with an external magnetic field $B_z=1$T
along $z$ axis the ground state manifold is given by the spin
aligned states
\begin{equation}\ket{T_+}=\hat{e}^\dagger_{L\uparrow}\hat{e}^\dagger_{R\uparrow}\ket{vac}=\ket{\uparrow\uparrow}\nonumber\end{equation}
and
\begin{equation}\ket{T_-}=\hat{e}^\dagger_{L\downarrow}\hat{e}^\dagger_{R\downarrow}\ket{vac}=\ket{\downarrow\downarrow},\nonumber\end{equation}
and the spin-anti-aligned states
\begin{equation}\ket{T_0}=\frac{1}{\sqrt{2}}\big(\hat{e}^\dagger_{L\uparrow}\hat{e}^\dagger_{R\downarrow}
+\hat{e}^\dagger_{L\downarrow}\hat{e}^\dagger_{R\uparrow}\big)\ket{vac}
=\frac{1}{\sqrt{2}}\big(\ket{\downarrow\uparrow}+\ket{\downarrow\uparrow}\big)\nonumber\end{equation}
and
\begin{equation}\ket{(1,1)S}=\frac{1}{\sqrt{2}}\big(\hat{e}^\dagger_{L\uparrow}\hat{e}^\dagger_{R\downarrow}
-\hat{e}^\dagger_{L\downarrow}\hat{e}^\dagger_{R\uparrow}\big)\ket{vac}
=\frac{1}{\sqrt{2}}\big(\ket{\downarrow\uparrow}-\ket{\downarrow\uparrow}\big)\nonumber\end{equation}
with energy gaps due to the Zeeman splitting and exchange energy
shown in Fig.~\ref{fig:energylevel}. The notation $(n_L,n_R)$ labels
the number of electrons in the left and right quantum dots. The
doubly occupied state $\ket{(0,2)S}$ is coupled via tunneling $T$ to
the singlet state $\ket{(1,1)S}$. The double-dot system can be
described by an extended Hubbard Hamiltonian
\begin{align}
&\hat{H}=(E_\text{os}+\mu)\sum_{i,\sigma}\hat{n}_{i,\sigma}-T\sum_{\sigma}\big(\hat{c}^\dagger_{L,\sigma}\hat{c}_{R,\sigma}+\text{hc}\big)\\&
+U\sum_i\hat{n}_{i,\uparrow}\hat{n}_{i,\downarrow}+W\sum_{\sigma,\sigma'}\hat{n}_{L,\sigma}\hat{n}_{R,\sigma'}+\Delta\sum_{\sigma}\big(\hat{n}_{L,\sigma}-\hat{n}_{R,\sigma}\big)\nonumber
\end{align}
for $\hat{c}_{i,\sigma}$ ($\hat{c}^\dagger_{i,\sigma}$) annihilating
(creating) an electron in quantum dot~$i\in\{L,R\}$ with
spin~$\sigma\in\{\uparrow,\downarrow\}$,
$\hat{n}_{i,\sigma}=\hat{c}^\dagger_{i,\sigma}\hat{c}_{i,\sigma}$ a
number operator, and $\Delta$ an energy offset yielded by the
external electric field along $x$ axis shown in
Fig.~\ref{fig:schematic}. The first term corresponds to on-site
energy~$E_\text{os}$ plus site-dependent field-induced
corrections~$\mu$. The second term accounts for $i \leftrightarrow
j$ electron tunneling with rate~$T$, the third term is the on-site
charging cost~$U$ to put two electrons with opposite spin in the
same dot, and the fourth term corresponds to inter-site Coulomb
repulsion, and the forth term is the cost of putting one electron at
site $i$ and another electron at site $j$.

In the basis $\{\ket{(1,1)S},\ket{(0,2)S}\}$, the Hamiltonian can be
deduced as
\begin{equation}
\hat{H}_\text{d}=-\Delta\ket{(0,2)S}\bra{(0,2)S}+T\ket{(1,1)S}\bra{(0,2)S}+\text{hc}.
\label{eq:deduceHam1}
\end{equation}
With the energy offset $\Delta$, degenerate perturbation theory in
the tunneling $T$ reveals an avoided crossing at this balanced point
between $\ket{(1,1)S}$ and $\ket{(0,2)S}$ with an energy gap
$\omega=\sqrt{\Delta^2+4T^2}$, and the effective tunneling between
the left and right dots with the biased energies $\Delta$ is changed
from $T$ to $\omega/2$.

We choose the singlet state and one of the triplet states as our
qubit:
\begin{equation}
\ket{0}\equiv\ket{(1,1)S},\text{ }\ket{1}\equiv\ket{T_0},
\label{eq:qubit}
\end{equation}
and the doubly occupied state as an ancillary state
\begin{equation}
\ket{a}\equiv \ket{(0,2)S}.
\end{equation}
The essential idea is to use an effective electric dipole moment
associated with singlet states $\ket{0}$ and $\ket{a}$ of a NW DQD
coupled to the oscillating voltage associated with a TLR shown in
Fig.~\ref{fig:schematic}. We consider a TLR with length $L$, the
capacitance per unit length $C_0$ and the characteristic impedance
$Z_0$. A capacitive coupling $C_c$ between the NW DQD and TLR causes
the electron charge state to interact with excitations in the
transmission line. We assume that the dot is much smaller than the
wavelength of the resonator excitation, so the interaction strength
can be derived from the electrostatic potential energy of the system
\begin{equation}
\hat{H}_\text{int}=e\hat{V}v\ket{a}\bra{a},
\end{equation}
where $e$ is the electron charge,
\begin{equation}
v=\frac{C_c}{C_\text{tot}}, \text{
}\hat{V}=\sum_{n}\sqrt{\frac{\hbar\omega_n}{LC_0}}\big(\hat{a}_n+\hat{a}_n^\dagger\big)
\end{equation}
is the voltage on the TLR near the left dots,
{$\hat{a}_n,\hat{a}^\dagger_n$} are the creation and annihilation
operators for the mode $k_n=[(n+1)\pi]/L$ of the TLR, and
$C_\text{tot}$ is the total capacitance of the DQD. The fundamental
mode frequency of the TLR is $\omega_0=\pi/LZ_0C_0$. The TLR is
coupled to a capacitor $C_e$ for writing and reading the signals.
Neglecting the higher modes of the TLR and working in the rotating
frame with the rotating wave approximation, we obtain an effective
Hamiltonian as
\begin{equation}
\hat{H}_\text{eff}=\omega_0\hat{a}^\dagger\hat{a}+\omega\ket{a}\bra{a}+g\big(\hat{a}\ket{a}\bra{0}+\text{hc}\big)
\label{eq:intHam}
\end{equation}
with $\hat{a}$ ($\hat{a}^\dagger$) the annihilation (creation)
operator of the resonator field, and the effective coupling
coefficient
\begin{equation}
g=\frac{1}{2}e\frac{C_\text{c}}{LC_\text{tot}C_0}\sqrt{\frac{\pi}{Z_0}}\sin2\theta
\label{eq:g}
\end{equation}
with $\theta=\frac{1}{2}\tan^{-1}(\frac{2T}{\Delta})$.

The interaction between the TLR and qubit states is switchable via
tuning the electric field along $x$ axis. In the case of the energy
offset yielded by the electric field $\Delta\approx 0$, we obtain
the maximum value of the coupling between the TLR and singlets in
DQDs. That is so-called the optimal point. Whereas $\Delta\gg T$,
$\theta$ tends to $0$, the interaction is switched off.

The transition from $\ket{1}$ to $\ket{a}$ is driven by a classical
laser field with a Rabi frequency $\Omega$. The interaction
Hamiltonian is given by
\begin{equation}
\hat{H}_\text{dr}=\Omega
e^{-i\omega_\text{dr}t}\ket{a}\bra{1}+\Omega
e^{i\omega_\text{dr}t}\ket{1}\bra{a}, \label{eq:dr}
\end{equation}where $\omega_\text{dr}$ is the drive frequency.

\subsection{Physical realization}

A realization of DQDs defined using local gates to electrostatically
deplete InAs NWs grown by chemical beam epitaxy was
reported~\cite{Fasth05}. The quantum-mechanical tunneling $T$
between the two quantum dots is about $0-150\mu$eV~\cite{Fasth05}.
Thus at the optimal point $\Delta\approx 0$ where the coupling is
strongest, the energy gap between the singlets is about
$\omega\sim2T\simeq 0-72$GHz. A small-diameter ($\sim65$nm),
long-length ($\sim270$nm) and $g^*=-13$~\cite{MT05} InAs NW is
positioned perpendicularly to the transmission line and containing
DQDs that are elongated along the NW shown in
Fig.~\ref{fig:schematic}. The external magnetic field along $z$ axis
is about $B_z=1$T to make sure that the energy splitting
$E_z=g^*\mu_B B_z$ between the two triplet states $\ket{T_{\pm}}$ is
larger than $\omega$.

The TLR can be fabricated with existing lithography
techniques~\cite{Wallraff04}. The dots can be placed within the TLR
formed by the transmission line to strongly suppress the spontaneous
emission.  To prevent a current flow, the NW and transmission line
need to be separated by some insulating coating material obtained
for example by atomic layer deposition. We assume that the TLR is
$3$cm long and $10\mu$m wide, $Z_0=50\Omega$ which implies for the
fundamental mode $\omega_0=\pi/LC_0Z_0=2\pi\times10$GHz. In
practice, careful fabrication permits a strong coupling capacitance,
with $C_\text{tot}\approx 5.1C_\text{c}$~\cite{Fasth05}, so that the
coupling coefficient $g\sim 2\pi\times120$MHz is achievable due to
the numerical estimations in Eq.~(\ref{eq:g}). The frequency
$\omega_0$ and coupling coefficient $g$ can be tuned via $LC_0$.
With a magnetic field about $1$T the resonators in coplanar
waveguides with $Q\sim 10^3-10^4$ have already been demonstrated
in~\cite{Frunzio05}.

The effect of photon assisted tunneling (PAT) in our system is
harmful because it destroys the qubit by lifting spin-blockade. To
avoid this, our strategy is to close enough the tunneling barriers
to the leads.

\section{universal quantum computing}
\subsection{Single-qubit gate operations}

First we consider the zero-detuning case in which the fundamental
mode frequency of the TLR is $\omega_0\approx \omega$. The
Hamiltonian (\ref{eq:intHam}) has the same form as the
Jaynes-Cummings Hamiltonian of a two-level system with a single-mode
resonator field. In the case when the TLR is initially in the photon
number state $\ket{n}_\text{r}$, the time evolution of the system,
governed by the Hamiltonian (\ref{eq:intHam}), is described by
\begin{align}
&\ket{0}\ket{n}_\text{r}\rightarrow\cos\sqrt{n}gt\ket{0}\ket{n}_\text{r}-i\sin\sqrt{n}gt\ket{a}\ket{n-1}_\text{r},\\
&\ket{a}\ket{n}_\text{r}\rightarrow
-i\sin\sqrt{n+1}gt\ket{0}\ket{n+1}_\text{r}+\cos\sqrt{n+1}gt\ket{a}\ket{n}_\text{r}.\nonumber
\end{align}

From the Hamiltonian of a drive on the DQDs shown in (\ref{eq:dr}),
it is straightforward to see that a pulse of duration $t$ results in
the following rotation:
\begin{align}
&\ket{1}\rightarrow \cos\frac{\Omega}{2}t\ket{1}-\sin\frac{\Omega}{2}t\ket{a},\nonumber\\
&\ket{a}\rightarrow
-i\sin\frac{\Omega}{2}t\ket{1}+\cos\frac{\Omega}{2}t\ket{a}.
\end{align}

It has been reported in~\cite{Zheng05} that the structure of the TLR
and qubits tuned on-resonance can be used to implement an entangling
gate on spin qubits of NW DQDs via the adiabatic evolution of the
dark states.

\begin{figure}[tbp]
   \includegraphics[width=8.5cm]{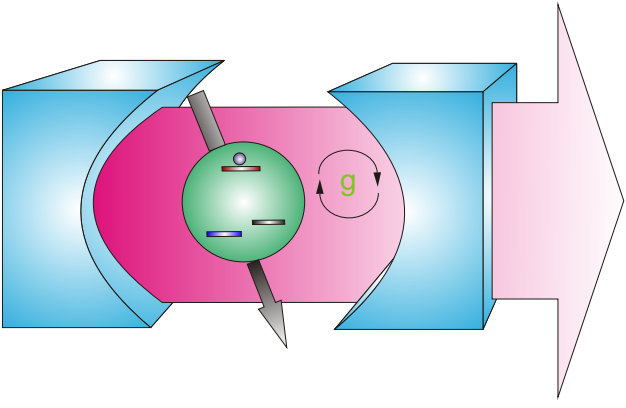}
  \caption{Standard representation of a cavity QED system,
  comprising a single mode of the electromagnetic field in a
  cavity coupled with  a strength $g$ to a three-level system
  and a classical drive.}
  \label{fig:cqed}
\end{figure}

In this paper, we show a different proposal and consider the case
where the TLR and qubits are tuned off resonant, which leads to
lifetime enhancement of the qubits and implements coherent control.
Assume that the classical field and TLR are detuned from the
transitions by $\delta_1=\omega_\text{dr}-(\omega-J)$ and
$\delta_2=\omega_0-\omega$, respectively, the Hamiltonian for single
DQD coupled to the TLR and driven by a classical field is
\begin{equation}
\hat{H}_\text{1q}=\omega_0\hat{a}^\dagger\hat{a}+\omega\ket{a}\bra{a}+g\big(\hat{a}\ket{a}\bra{0}+\Omega
e^{-i\omega_\text{dr}t}\ket{a}\bra{1}+\text{hc}\big)
\end{equation}

If $\delta_1,\delta_2\gg \Omega,g$ is satisfied, the upper level
$\ket{a}$ can be adiabatically eliminated. We then obtain the
effective Hamiltonian of the system as
\begin{equation}
\hat{H}^\text{eff}_\text{1q}=\frac{\Omega^2}{\delta_1}\ket{1}\bra{1}
+\frac{g^2}{\delta_2}\hat{a}^\dagger\hat{a}\ket{0}\bra{0}+\lambda\big(\hat{a}\ket{1}\bra{0}+\text{hc}\big),
\label{eq:1q}
\end{equation}
where $\lambda=\Omega g/2(\delta_1+\delta_2)$. The first two terms
describe the Stark shifts for the spin states $\ket{0}$ and
$\ket{1}$, induced by the classical field and resonator mode,
respectively. The last term is the Raman coupling of the two spin
states.

For single spin qubits in $\{\ket{0},\ket{1}\}$ coupled with
effective strength $\lambda$ to the TLR, driven by a classical field
which is detuned from the TLR, the Hamiltonian (\ref{eq:1q}) can be
used to implement single qubit rotations along $x$ axis via the Rabi
oscillation between the states $\ket{0}$ and $\ket{1}$ shown in
Fig.~\ref{fig:cqed}.

\subsection{Two-qubit gate operations}
Now we consider there are two spin qubits coupled to the TLR. From
the Hamiltonian (\ref{eq:1q}), in the case of
$\delta_2-\delta_1\gg\lambda$, there is no energy exchange between
the DQD system and TLR. The energy conversing transitions are
between $\ket{1_10_2}\ket{n}_\text{r}$ and
$\ket{0_11_2}\ket{n}_\text{r}$. The effective Rabi frequency for the
transitions between these states, mediated by
$\ket{0_10_2}\ket{n+1}_\text{r}$ and
$\ket{1_11_2}\ket{n-1}_\text{r}$ is given by
\begin{align}
\lambda'=&\frac{\bra{1_10_2n}\hat{H}_\text{tot}\ket{0_10_2n+1}\bra{1_10_2n+1}\hat{H}_\text{tot}\ket{0_11_2n}}{\delta_2-\delta_1}\nonumber\\
&+\frac{\bra{1_10_2n}\hat{H}_\text{tot}\ket{1_11_2n-1}\bra{1_11_2n-1}\hat{H}_\text{tot}\ket{0_11_2n}}{-(\delta_2-\delta_1)}\nonumber\\
&=\frac{\lambda^2}{\delta_2-\delta_1},
\end{align} where
$\hat{H}_\text{tot}=\sum_{j=1,2}\hat{H}_\text{1q}^j$. The effective
Hamiltonian for two qubits turns to be
\begin{align}
\hat{H}_\text{2q}=&\sum_{j=i,2}\frac{g^2}{\delta_2}\hat{a}^\dagger\hat{a}\ket{0}_j\bra{0}
+\frac{\Omega^2}{\delta_1}\ket{1}_j\bra{1}+\lambda'\hat{a}\hat{a}^\dagger\big(\ket{1}_j\bra{1}\nonumber\\
&-\ket{0}_j\bra{0}\big)+\lambda'\big(\ket{1}_j\bra{0}\otimes\ket{0}_j\bra{1}+\text{hc}\big).
\label{eq:2q}
\end{align}
The third and forth terms are the photon-number dependent Stark
shifts induced by the Raman transition, and the last term is the
induced dipole coupling between the two spin qubits. If the
resonator mode is initially in the vacuum state it will remain in
the vacuum state throughout the process. Then the effective
Hamiltonian for the two qubits is reduced to
\begin{equation}
\hat{H}^\text{eff}_\text{2q}=\sum_{j=1,2}\frac{\Omega^2}{\delta_1}\ket{1}_j\bra{1}+\lambda'\big(\hat{\sigma}_+^1\hat{\sigma}_-^2+\text{hc}\big),
\label{eq:2qr}
\end{equation}
where $\hat{\sigma}_+=\ket{1}\bra{0}$ and
$\hat{\sigma}_-=\ket{0}\bra{1}$.

The evolution of the effective two-qubit Hamiltonian can be used to
implement an entangling two-qubit gate---$\sqrt{i\text{SWAP}}$. In a
frame rotating at the qubit's frequency, the Hamiltonian
(\ref{eq:2qr}) generates the evolution
\begin{align}
U_{2q}(t)=&\prod_{j=1,2}\text{exp}\left[-i\frac{\Omega^2}{\delta_1}t\ket{1}_j\bra{1}\right]\nonumber\\
&\times \left(
         \begin{array}{cccc}
           1 & 0 & 0 & 0 \\
           0 & \cos\lambda' t & i\sin\lambda' t & 0 \\
           0 & i\sin\lambda' t & \cos\lambda' t & 0 \\
           0 & 0 & 0 & 1 \\
         \end{array}.
       \right)
\label{eq:u2q}
\end{align}
Up to the phase factor, it corresponds at
$t_\text{2q}=\pi/4\lambda'$ to a $\sqrt{i\text{SWAP}}$ logical
operation. Up to single-qubit gates, the above operation is
equivalent to the controlled-not gate. Together with single-qubit
gates, the interaction $\hat{H}^\text{eff}_\text{2q}$ is therefore
sufficient for universal quantum computing.

When the qubits are detuned form each other, the off-diagonal
coupling provided by $\hat{H}^\text{eff}_\text{2q}$ is only weakly
effective and the coupling is for all practical purposes turned off.
Two-qubit logical gates in this setup can therefore be controlled by
individually tuning the qubits. Moreover, single- and two-qubit
logical operations on different qubits and pairs of qubits can both
be realized simultaneously, a requirement of reach presently known
thresholds for fault-tolerant quantum computation~\cite{Aharonov96}.


Hence we have built a universal set of gates for quantum computing
with semiconductor DQDs coupled to a resonator field. The
feasibility of single-qubit gates has already been proved
in~\cite{WSB+05} experimentally. For the two-qubit gate, we realize
it with the off-resonant interaction between both qubits and TLR.
With the experimental parameters
$\{T,J,\omega_0,\omega,\omega_\text{dr},g,\Omega\}/2\pi=\{2.5,0.25,10,5,9.74,0.12,1\}$GHz,
the detunings $\delta_1/2\pi=4.99$GHz and $\delta_2/2\pi=5$GHz, and
the efficient coupling coefficients $\lambda=3$MHz and
$\lambda'=0.9$MHz, we can estimate the time scaling for quantum
computing. The operating time for the single-qubit rotation along
$x$ axis is $t_x\sim1/\lambda\approx300$ns with the above
parameters. The single-qubit rotation along $z$ axis takes the same
time scaling as the two-qubit gate in (\ref{eq:2qr}) $t_\text{2q}$
which satisfies $\lambda' t_\text{2q}=\pi/4$ and is calculated as
$t_\text{2q}\approx1\mu$s.

\subsection{Fidelity of two-qubit gates}

\begin{figure}[tbp]
   \includegraphics[width=8.5cm]{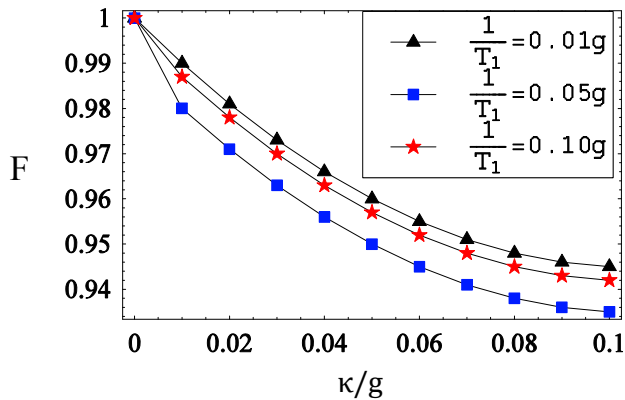}
  \caption{(color online). Fidelity ($F$) of the two-qubit $\sqrt{i\text{SWAP}}$
  gate vs the resonator decay rate $\kappa$ with the experimental parameters
  $\{\omega_0,\omega,\omega_\text{dr},g,\Omega\}/2\pi=\{10,5,9.74,0.12,1\}$GHz. The triangled, stared, boxed lines
  describe the cases of the spontaneous emission rate of the singlet states
  $1/T_{1}=0.01g,\text{ }0.05g,\text{ }0.1g$, respectively.}
   \label{fig:fidelity}
\end{figure}

Now we analyze the effect on gate operations due to noise and derive
the fidelity of two-qubit gates. We use the two-qubit gate
Eq.~(\ref{eq:u2q}) as an example. With the time dependent
fluctuations $\delta \lambda'(t)$ of the effective coupling
coefficient $\lambda'$, the evolution operator of the system becomes
\begin{align}
U'_{2q}=U_{2q}\exp\big[-\text{i}\int_{0}^{t_\text{2q}}\text{d}t\delta\lambda'(t)(\hat{\sigma}_+^1\hat{\sigma}_-^2+\hat{\sigma}_-^1\hat{\sigma}_+^2)\big],
\end{align}
where the unwanted phase
$\phi=\int_{0}^{t_\text{2q}}\text{d}t\delta\lambda'(t)$. The
distribution of the unwanted phase becomes Gaussian distribution
because $\lambda'$ is in Gaussian distribution. With the parameters
above, we numerically calculate the variances of the unwanted phase
$\text{Var}(\phi)\sim2\times10^{-3}\pi$.

Furthermore, the decoherence such as dephasing reduces the fedility
of gate operations as well. We analyze the dephasing rate due to the
variations of the AC Stark shift
$g^2/\delta_2\hat{a}^\dagger\hat{a}\hat{\sigma}_z$ (where
$\hat{\sigma}_z=\ket{0}\bra{0}-\ket{1}\bra{1}$) caused by quantum
fluctuations of the number of photon $\bar{n}$ within the resonator.
To determine the dephasing rate, we assume that the resonator is
driven at the bare resonator frequency $\omega_0$ and the pull of
the resonance is small compared to the linewidth $\kappa$. The
relative phase accumulated between the two singlet states $\ket{0}$
and $\ket{a}$ is
$\vartheta(t)=2\frac{g^2}{\delta_2}\int_0^t\text{d}t'n(t')$, which
yields a mean phase advance
$\langle\vartheta\rangle=2g^2\bar{n}t/\delta_2$. Dephasing can be
evaluated by the decay of the correlator
$\Big\langle\exp\left[i\int_0^t\text{d}t'\vartheta(t')\right]\Big\rangle$.
If the resonator is not driven the photon number correlator rather
decays at a rate $\kappa$ and the rate of transimission on-resonance
is $\gamma_\vartheta=\bar{n}\kappa/2$. In the dispersive regime, the
dephasing rate is reduced to
$\gamma_\vartheta=8\bar{n}(\frac{g^2}{\delta_2})^2\frac{1}{\kappa}$.

From the analysis, we show that even the decoherence and noise occur
over the gate operation, we can still implement a universal set of
gates with high fidelities. In Fig.~\ref{fig:fidelity}, we show the
fidelity $F=_\text{int}\bra{\varphi}U_{2q}^\dagger \rho
U_{2q}\ket{\varphi}_\text{int}$ of the two-qubit gate
$\sqrt{i\text{SWAP}}$ as a function of the resonator decay $\kappa$
and spontaneous emission of the DQD singlet state $1/T_1$, where
$\rho$ is the reduced density matrix calculated by solving the
master equation with decoherence and tracing out the resonator
photon, and $\ket{\varphi}_\text{int}$ is the initial states of the
spin states. With the experimental parameters
$\{\omega_0,\omega,\omega_\text{dr},g,\Omega,\kappa,1/T_1\}/2\pi=\{10,5,9.74,0.12,1,0.001,0.001\}$GHz,
the fidelity is achieved as high as $0.991$.

For single-qubit $\hat{\sigma}_x$ gate, the unwanted phase is
$\int_0^{t_x}\text{d}t\delta\lambda(t)$. With the same method, we
can calculate the variance of the phases.

\subsection{Initialization and transportation}
Initialization of qubit states can be implemented by an adiabatic
passage between the two singlet states $\ket{0}$ and
$\ket{a}$~\cite{Taylor07}. Controllably changing $\Delta$ allows for
adiabatic passage to past the charge transition, with $\ket{a}$ as
the ground state if $\Delta \gg T$ achieved. First we turn on the
external electric field along $x$ axis and prepare the two electrons
of NW DQDs in the state $\ket{a}$ by a large energy offset $\Delta$.
We change $\theta$ in Eq.~(\ref{eq:g}) adiabatically to $\pi/4$ by
tuning the electric field, and then initialize the qubits in the
state $\ket{0}$.

The SWAP operation~\cite{Taylor06}, where a qubit state is swapped
with a photonic state of the TLR, can be used to implement
transmission of qubits. If there is no photon in the TLR, with the
evolution time $\pi/\lambda$, a qubit is mapped to the photonic
state in the TLR
\begin{equation}
\big(\alpha\ket{0}+\beta\ket{1}\big)\ket{0}_\text{r}\longrightarrow
\ket{0}\big(\alpha\ket{0}+\beta\ket{1}\big)_\text{r}.
\end{equation}
Then we switch off the coupling between this qubit and TLR and
switch on that between the desired qubit and TLR via the local
electric fields along $x$ axis. After the same evolution time, the
previous qubit state is transmitted to the desired qubit via the
interaction with the TLR. The time for transmitting a qubit to a
photonic qubit in the TLR is about
$t_\text{tr}=\pi/\lambda\approx900$ns with the experimental
parameters shown in Sec.~III.

\subsection{Readout}

To perform a measurement of qubits, the classical field and TLR are
tuned from the respective transitions modeled by Eq. (\ref{eq:1q}).
In the dispersive regime ($\delta_1,\delta_2\gg \Omega,g$), the
energy gap between the dressed states $\ket{0}_\text{r}$ and
$\ket{1}_\text{r}$ is $\omega_0-g^2/\delta_2$ for the qubit in the
state $\ket{0}$, while the energy gap $\omega_0$ for the state
$\ket{1}$ remains unchanged. The operator being probed is
$\hat{\sigma}_z$ and the qubit-measurement apparatus interaction
Hamiltonian is $g^2/\delta_2\hat{a}^\dagger\hat{a}\hat{\sigma}_z$,
such that
$\left[\hat{\sigma}_z,g^2/\delta_2\hat{a}^\dagger\hat{a}\hat{\sigma}_z\right]=0$.
Depending on the qubit being in the states $\ket{0}$ or $\ket{1}$
the transmission spectrum presents a peak of width $\kappa$ (the
resonator decay rate) at $\omega_0-g^2/\delta_2$ or $\omega_0$. This
dispersive pull of the resonator frequency is $0\sim
g^2/\kappa\delta_2$, and the pull is power dependent and decreases
in magnitude for photon numbers inside the TLR~\cite{Blais04}. Via
microwave irradiation of the TLR by probing the transmitted or
reflected photons, the readout of qubits can be realized and
completed on a time scaling $t_\text{m}=1/\gamma_\vartheta$, where
$\gamma_\vartheta$ is the dephasing rate due to quantum fluctuations
of the number of photon $\bar{n}$ within the TLR shown in Sce.~III
C. Compared to the dephasing rate of transmission on resonance
$\gamma_\vartheta=\bar{n}\kappa/2$, in the dispersive regime the
phase noise induced by the AC Stark shift
$g^2/\delta_2\hat{a}^\dagger\hat{a}\hat{\sigma}_z$ results in the
dephasing rate
$\gamma_\vartheta=8\bar{n}(\frac{g^2}{\delta_2})^2\frac{1}{\kappa}$
and an enhanced lifetime of spin qubits. This approach can serve as
a high efficiency quantum nondemolition dispersive readout of the
qubit states: $P_1=\ket{0}\bra{0};\ P_2=\ket{1}\bra{1}$. Readout of
qubits takes the time $t_\text{m}\approx1.5$ns in the case $\bar
n=100$ with the experimental parameters shown above.

\subsection{Decoherence}

In Sec.~III C, we show the decoherence occurring over the two-qubit
gate operation, now we analyze the dominant noise sources of the
system existing during all precessings, which include the spin phase
noise due to hyperfine coupling, the charge-based dephasing and
relaxation occurring during gate operations and transportation, and
the photon loss due to the resonator decay. The characteristic
charge dephasing with a rate $T_2^{-1}$. The time-ensemble-averaged
dephasing time $T_2^*$ is limited by hyperfine interactions with
nuclear spins. Coupling to a phonon bath causes relaxation of the
charge system in a time $T_1$. The decay of the TLR $\kappa$ is
considered as another dominant source of decoherence.

The hyperfine interactions with the gallium arsenide host nuclei
causes nuclear spin-related dephasing $T_2^*$. The hyperfine field
can be treated as a static quantity, because the evolution of the
random hyperfine field is several orders slower than the electron
spin dephasing. In the operating point, the most important
decoherence due to hyperfine field is the dephasing between the
singlet state $\ket{(1,1)S}$ and one of the triplet state
$\ket{T_0}$. By suppressing nuclear spin fluctuation, the dephasing
time can be obtained by quasi-static approximation as
$T_2^*=1/g\mu_B\langle\Delta B_n^z\rangle_\text{rms}$, where $\Delta
B_n^z$ is the nuclear hyperfine gradient field between two coupled
dots and rms means a root-mean-square time-ensemble average. A
measurement of the dephasing time $T_2^*\sim4$ns was demonstrated
in~\cite{Trif08} and we expect coherently driving the qubit well
prolong the $T_2^*$ time up to $1\mu$s and with echo up to
$10\mu$s~\cite{Petta05}.

For the charge relaxation time $T_1$, the decay is caused by
coupling qubits to a phonon bath. With the spin-boson model, the
perturbation theory gives an overall error rate from the relaxation
and incoherent excitation, with which one can estimate the
relaxation time $T_1\sim1\mu$s~\cite{Taylor06} which is studied in
great detail for the GaAs quantum dot in 2DEG and similar rate is
expected for NW quantum dots.

The charge dephasing $T_2$ rises from variations of the energy
offset $\Delta(t)=\Delta+\epsilon(t)$ with
$\langle\epsilon(t)\epsilon(t')\rangle=\int d\omega
S(\omega)e^{\text{i}\omega(t-t')}$, which is caused by the low
frequency fluctuation of the electric field. The gate bias of the
qubit drifts randomly when an electron tunnels between the metallic
electrode. Due to the low frequency property, the effect of the
$1/f$ noise on the qubit is dephasing rather than relaxation. At the
zero derivative point, compared to a bare dephasing time
$T_b=1/\sqrt{\int d\omega S(\omega)}$, the charge dephasing is
$T_2\sim \omega T_b^2$ near the optimal point $\Delta=0$. The bare
dephasing time $T_b\sim 1$ns was observed in~\cite{Hayashi04}. Then
the charge dephasing is estimated as $T_2\sim10-100$ns. Using
quantum control techniques, such as better high- and low-frequency
filtering of electronic noise, $T_b$ exceeding $1\mu$s was observed
in 2DEG~\cite{Petta05} (we assume a similar result for the present
case), which suppresses the charge dephasing.

The quality factor $Q$ of the TLR in the microwave domain can be
achieved $10^6$~\cite{Wallraff04}. In practice, the local external
magnetic field $1$T reduces the limit of the quality factor to
$Q\sim 10^3-10^4$~\cite{Frunzio05}. The dissipation of the TLR
$\kappa=\omega_0/Q$ leads to the decay time about
$100\text{ns}-1\mu\text{s}$ with the parameters
$\omega_0=2\pi\times10$GHz.

Thus, the operating times of all these gates
$\{t_x,t_\text{2q},t_\text{tr},t_\text{m}\}=\{300\text{ns},1\mu\text{s},900\text{ns},1.5\text{ns}\}$
are less than the minimum decoherence time.

\section{Summary}

Advances in fabrication have led to the development of solid-state
systems, with obvious potential for quantum computing. The
Heisenberg exchange coupling, optical dipole-dipole interactions,
capacitive coupling, and optical cavity-mediate interactions between
spin and charge states can be used to realize controlled quantum
state operations. In this paper we focus on NW DQDs quantum computer
which would capitalize on chip fabrication technology and could be
hybridized with existing computers. We propose a realization of
cavity QED via electrically controlled semiconductor spins of NW
DQDs coupled to a microwave TLR on a chip. Combining the advantages
of spin and charge states and avoiding the weak points of both, we
propose a mechanism to achieve a scalable architecture for quantum
computing with NW DQDs inside a TLR, namely via resonator-assisted
interaction which leads to an efficient, strong coupling between the
resonator photon and effective electric dipole of DQDs. Thus we
encode the quantum information in spin states and the gate operation
is implemented via the charge dipole transition which is driven by a
resonator. Initialization of qubits can be realized with an
adiabatic passage. With the switchable coupling to the TLR, we can
implement a universal set of quantum gates on any qubit. Because of
the switchable coupling between the double-dot pairs and TLR, we can
apply this entangling gate on any two qubits without affecting
others, which is not trivial for implementing scalable quantum
computing and generating large entangled state. The fidelities of
the gates in our protocol are studied including all kinds of major
decoherence, with promising results for reasonably achievable
experimental parameters and these results demonstrate the
practicality by way of current experimental technologies. Our work
shows how an experiment can be performed under existing conditions
to demonstrate the first architecture for quantum computing for spin
qubits in quantum dots in the laboratory.

\begin{acknowledgements}
This work has been supported by National Natural Science Foundation
of China, Grant No. 11004029, Jiangsu Province Natural Science
Foundation, Grant No. BK2010422, and Southeast University Startup
fund.
\end{acknowledgements}


\begin{references}
\bibitem {Gro97} L. Grover, \prl {\bf 79}, 325 (1997).
\bibitem {Sho94} P.W. Shor, Proc. 35th Annual Symp. on Found. of Comp. Sci.
    (Los Alamitos, CA: IEEE Computer Society Press, 1994) 124.
\bibitem {LD97} D. Loss, D.P. DiVincenzo, \pra 57, 120 (1998).
\bibitem {Kan98} B.E. Kane, Nature {\bf 393}, 133 (1998).
\bibitem {ABW+07} S.E.S. Andresen et al., Nanolett. {\bf 7}, 2000 (2007).
\bibitem {WSB+05} A. Wallraff et al., \prl {\bf 95}, 060501 (2005).
\bibitem{Im99} A. Imamoglu et al., Phys. Rev. Lett. {\bf 83}, 4204 (1999).
\bibitem{Petta05} J.R. Petta et al., Science {\bf 309}, 2180 (2005).
\bibitem{Johnson05} A.C. Johnson, Nature {\bf 435}, 925 (2005).
\bibitem{Taylor07} J.M. Taylor et al., Phys. Rev. B {\bf 76}, 035315 (2007).
\bibitem{GI06} G. Burkard, A. Imamoglu, Phys. Rev. B {\bf 74}, 041307(R) (2006).
\bibitem{Taylor06} J.M. Taylor, M.D. Lukin, arXiv: cond-mat/0605144.
\bibitem{Guo09} Z.R. Lin et al., Phys. Rev. Lett. {\bf 101}, 230501 (2008).
\bibitem{Xue10} P. Xue, Phys. Lett. A {\bf 374}, 2601 (2010).
\bibitem{Trif08} M. Trif, V.N. Golovach and D. Loss, Phys. Rev. B
{\bf 77}, 045434 (2008).
\bibitem{DiVincenzo} D.P. DiVincenzo, Fortschr. Phys. {\bf 48}, 771 (2000).
\bibitem{Fasth05} C. Fasth, A. Fuhrer, M.T. Bj\"{o}rk and L. Samuelson, Nano Lett.
{\bf 5}, 1487 (2005).
\bibitem{MT05}M.T. Bj\"{o}rk et al., Phys. Rev. B {\bf 72}, 201307(R) (2005).
\bibitem{Wallraff04} A. Wallraff et al., Nature {\bf 431}, 162 (2004).
\bibitem{Frunzio05}L. Frunzio et al., Applied Superconductivity, IEEE Transactions on {\bf 15}, 860 (2005).
\bibitem{Zheng05} S.B. Zheng, Phys.Rev. Lett. {\bf 95}, 080502 (2005); P.
Xue, Chin. Phys. Lett. {\bf 27}, 060301 (2010).
\bibitem{Aharonov96}D. Aharonov and M. Ben-Or, in {\it Proceedings of the
37th Annual Symposium on Foundations of Computer Science} (IEEE
Computer Society Press, Los Alamitos, CA, 1996), p. 46.
\bibitem{Blais04} A. Blais et al., Phys. Rev. A {\bf 69}, 062320
(2004); J. Gambetta et al., Phys. Rev. A {\bf74}, 042318 (2006); A.
Blais et al., Phys. Rev. A {\bf75}, 032329 (2007).
\bibitem{Hayashi04} T. Hayashi et al., Phys. Rev. Lett. {\bf 91}, 226804 (2003); J.R. Petta et al., {\it ibid} {\bf 93}, 186802 (2004).
\end{references}
\end{document}